\documentclass[conference]{IEEEtran}

\usepackage{graphicx}
\usepackage{cite,amsmath,amsfonts,graphicx}
\usepackage{bm}
\usepackage{amsbsy}
\usepackage{color}
\usepackage{soul}
\usepackage[caption=false]{subfig}

\begin{document}

\title{Microgrid Optimal State Estimation Over \\IoT Wireless Sensor Networks With \\Event-Based Measurements}

\author{\IEEEauthorblockN{Seyed~Amir~Alavi,~\IEEEmembership{Graduate Student Member,~IEEE,}			
            Mehrnaz~Javadipour,~\IEEEmembership{Student Member,~IEEE,}
            Kamyar~Mehran~\IEEEmembership{Member,~IEEE}}
\IEEEauthorblockA{School of Electronic Engineering and Computer Science\\
Queen Mary University of London\\
London E1 4NS, UK\\
\{s.alavi, m.javadipour, k.mehran\}@qmul.ac.uk }}

\maketitle

\begin{abstract}
In a microgrid, real-time state estimation has always been a challenge due to several factors such as the complexity of computations, constraints of the communication network and low inertia. In this paper, a real-time event-based optimal linear state estimator is introduced, which uses the send-on-delta data collection approach over wireless sensors networks and exhibits low computation and communication resources cost. By employing the send-on-delta event-based measurement strategy, the burden over the wireless sensor network is reduced due to the transmission of events only when there is a significant variation in the signals. The state estimator structure is developed based on the linear Kalman filter with the additional steps for the centralized fusion of events data and optimal reconstruction of signals by projection onto convex sets. Also for the practical feasibility analysis, this paper developed an Internet of things prototype platform based on LoRaWAN protocol that satisfies the requirements of the proposed state estimator in a microgrid.
\end{abstract}


\IEEEpeerreviewmaketitle

\section{Introduction}
Microgrids are small power systems that are able to operate independent of the main grid. The independent operation enables the optimal integration of renewable energy sources into the power system and also provides a higher degree of freedom in energy management comparing to the traditional power grid. Despite the advantages, microgrids poses low inertia, i.e. the system is more prone to instabilities driven by disturbances, and therefore, robust controllers should be employed to guarantee the continuous operation \cite{Alavi2018a, Parhizi2015}.

State estimation is an important part of a robust controller, as a high number of the robust control techniques are based on state feedback \cite{Qin2019, Alavi2019}. Also for the systems which are based on the output signal feedback, state estimation is inevitable for internal stability analysis and situational awareness (SA) \cite{Ghahremani2016, Amini2019}. The low inertia characteristic of the microgrids necessitates that the state estimator to work in real-time with a reasonable communication and computation cost \cite{Alavi2018}.

In the literature of state estimators, two different approaches have been taken, distributed and centralized state estimation \cite{Primadianto2017}. Both approaches have advantages that suits them for the specific application. Distributed state estimation approach is mainly used when the system is large and the computation cost of a centralized estimator would make the solution infeasible. Although distributed approaches remove the single point of failure problem, it requires a high number of computing agents for state estimation tasks, which is not appropriate for small to medium sized microgrids \cite{Spanos2005DynamicCF, Zohaib2018, Amini2019a}.

In contrast, the basic assumption in the centralized state estimation approach, is to have a single estimator, which collects the data from the sensors installed throughout the microgrid. Therefore, the sensors doesn't need to be smart and computationally powerful, as they only have to measure and send the data to the collector \cite{Abessi2016, Sakurai2012}. Furthermore, the rate at which the sensors transmit the measurements greatly affects the performance of the state estimator, the network traffic, and the energy consumption of battery based nodes \cite{Anta2010}. Traditional state estimator theories were originally developed based on the fact that the signals were sampled periodically with a predetermined sampling period. With the advent of Internet of things (IoT) communication technologies, this basic assumption is not practical anymore, because the IoT technology trend is moving toward lower speed communication for longer distances and reduced power for wireless transmission \cite{Alavi2018, Gungor2011}. Therefore, modernized state estimators should be designed as such that are able to fuse the event-based data from different sensors across the microgrid.

To address this need, the authors in this paper propose a centralized event-based optimal linear state estimator, suitable for medium sized microgrids, with Send-on-Delta (SoD) measurements. The estimator uses projection onto convex sets (POCS) technique \cite{Rzepka2018} to optimally reconstruct the sparse received data from the nodes and then reduces the estimation error of event-based Kalman filter. 

In Section \ref{data_model}, microgrid data modeling for both AC and DC ones is provided. Afterward the proposed estimator is introduced in Section \ref{architecture} based on the modeled data. The developed event-based Kalman estimator is formulated in Section \ref{kalman} and the POCS data recovery technique is discussed in Section \ref{reconstruction}. In Section \ref{prototype}, the implemented setup for evaluation of the estimation strategy is shown. Finally, Section \ref{results} presents the results of the analysis. The paper is concluded in Section \ref{conc}.

\section{Microgrid State Estimation Data Modeling}\label{data_model}
In this section the problem of microgrid state estimation is justified. An autonomous single bus microgrid consists of renewable energy sources (RESs), energy storage systems (ESSs), power electronic converters and loads. Two types of power systems can be used for the microgrid implementation, DC (Direct Current) and AC (Alternating Current) systems. Each of these systems are dynamic processes that can be modeled as a system of differential equations, either linear or non-linear. As any other types of dynamic systems, each process has inputs, outputs and the internal state variables. The set of measurements available for state estimation in this paper is assumed as:

{\center\textbf{AC microgrid}}
\begin{IEEEeqnarray}{lll}
v_i \in V,\ \text{RMS voltage of buses} \IEEEnonumber \\
p_i \in P,\ \text{Active power injected into each bus} \IEEEnonumber \\
q_i \in Q,\ \text{Reactive power injected into each bus}
\end{IEEEeqnarray}

{\center\textbf{DC microgrid}}
\begin{IEEEeqnarray}{lll}
v_i \in V,\ \text{RMS voltage of buses} \IEEEnonumber \\
i_i \in I,\ \text{Injected current into each bus}
\end{IEEEeqnarray}

For AC microgrids, other variables can be chosen such as phase of voltage, but as phasor measurement units are expensive and need high speed synchronization, indirect methods with active and reactive power are recommended and used.

\section{Proposed Estimation Architecture}\label{architecture}
\begin{figure}[!t]
\centering
\includegraphics[width=3.5in]{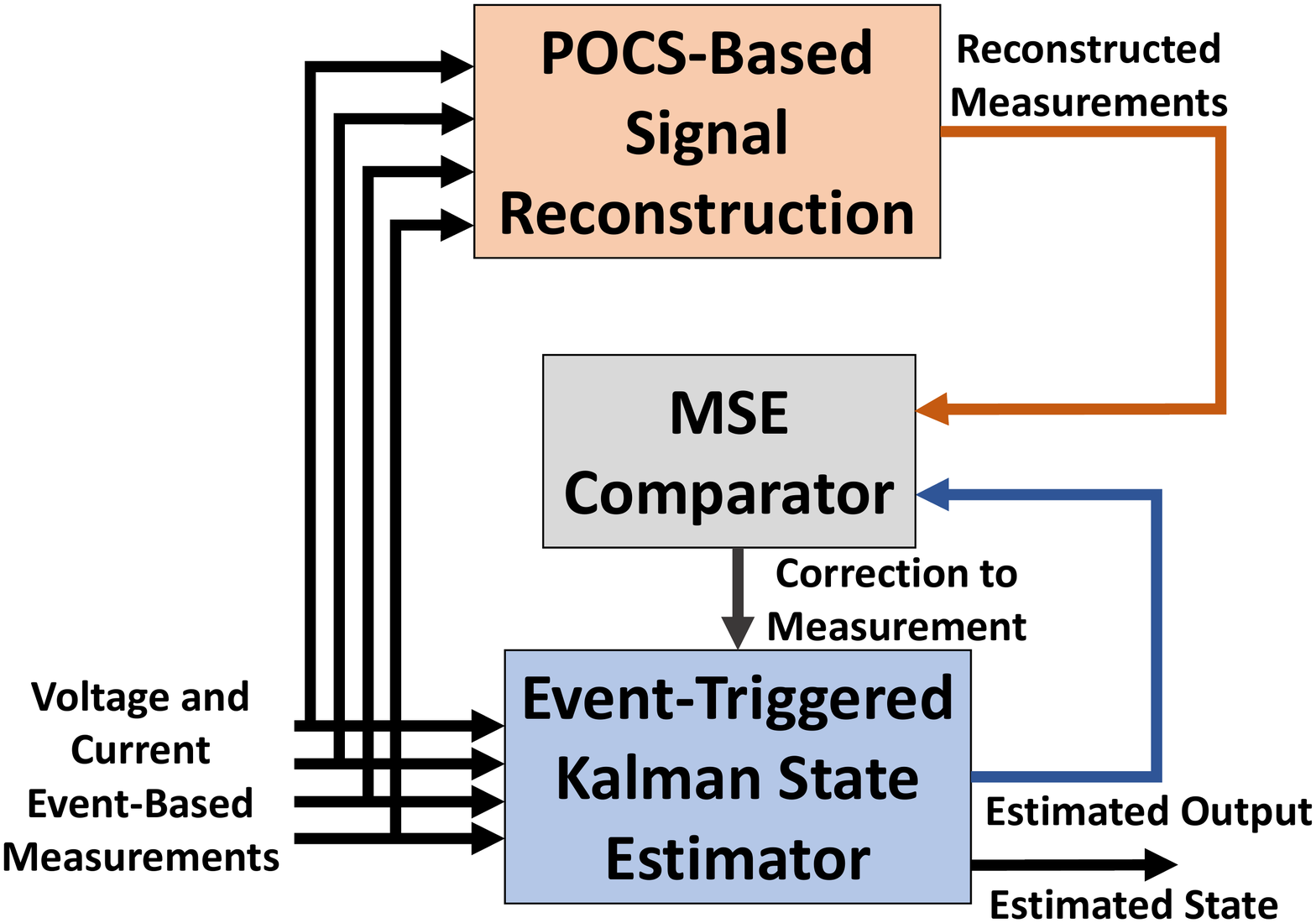}
\caption{Structure of the proposed event-based optimal state estimator.}
\label{estimator_structure}
\end{figure}

\figurename ~\ref{estimator_structure} shows the structure of the proposed event-based state estimator. It consists of three parts, event-based adaptive Kalman state estimator, event-based signal conditioner and the mean square error (MSE) comparator. The microgrid estimation input variables, defined in Section \ref{data_model}, are sampled with send-on-delta (SoD) measurement approach. The proposed event-based Kalman filter works based on the knowledge that the signals between each events are bounded by the $\delta$ threshold of SoD sampler. The proposed signal conditioner, reconstructs the original signal based on the events using the projection onto convex sets algorithm (POCS), which is mainly used in literature as a promising approach for low quality image reconstruction. Finally the MSE comparator, decides on updating the state estimator based on the error between reconstructed signal and the predicted output of Kalman filter.

\section{Event-Based Kalman Filter Design}\label{kalman}
By mapping the microgrid variables into the following dynamic system which is the state space realization the microgrid dynamics, the state estimation problem is formulated as:
\begin{IEEEeqnarray}{ll}
\dot{x}=Ax\left(t\right)+w\left(t\right)\IEEEnonumber \\ y(t)=Cx(t)+v(t)
\end{IEEEeqnarray}
where $x\mathrm{\ }\mathrm{\in}\mathrm{\ }{\mathrm{R}}^{\mathrm{n}}$ is the estimated state and $y\ \mathrm{\in}\mathrm{\ }{\mathrm{R}}^{\mathrm{p}}\mathrm{\ }$is the output measurement. The process noise $w\left(t\right)$ and measurement noise $v(t)$ are the uncorrelated, zero-mean white Gaussian random processes, satisfying the following assumptions:
\begin{IEEEeqnarray}{lll}
\text{E}\left\{w(t) \; w(s)'\right\} = Q \; \delta (t-s)
\\
\text{E}\left\{v(t) \; v(s)'\right\} = R \; \delta (t-s)
\\
\text{E}\left\{w_i\left(t\right){v_j\left(s\right)}'\right\} =0,\;\;1 \leq i \leq n, \;\;1 \leq j \leq p
\end{IEEEeqnarray}
where $w_i$ and $v_j$ are the \textit{i}-th and \textit{j}-th elements of the $w$ and $v$, respectively. Also, $R$ is the measurement noise covariance, and $Q$ is the process noise covariance. It is presumed that the\textit{ i}-th sensor only transmits the data when the difference between the current sensor value and the previously transmitted value is greater than ${\delta }_i$.

The states are also estimated periodically with the period of $T$. For simplicity, it is assumed that there is no delay in the sensor data transmission. Using the SoD method \cite{Li2016}, the estimator continuously samples the available data with a period of $T$ from the sensors. For example, if the last received $i$-th sensor value is $y_i$ at the time $t_{last,i}$, and there is no $i$-th sensor data received for ${t>t}_{last,i}$, then the estimator can estimate $y_i(t)$ as:
\begin{equation}
y_i\left(t_{last,i}\right)-{\delta }_i\le \ y_i\left(t\right)\le y_i\left(t_{last,i}\right)+{\delta }_i
\end{equation}

The last received \textit{i}-th sensor data is used to compute the output $y_{computed,i}$ even if there is no sensor data transmission:
\begin{equation}
\label{eq:2}
y_{computed,i}\left(t\right)=y_i\left(t_{last,i}\right)=C_ix\left(t\right)+v_i\left(t\right)+{\Delta }_i\left(t,t_{last,i}\right)
\end{equation}
where ${\mathrm{\Delta }}_i\left(t,t_{last,i}\right)\mathrm{=}y_i\left(t_{last,i}\right)\mathrm{-}y_i\left(t\right)$ and: 

\begin{equation}
\label{eq:3}
\left|{\Delta }_i\left(t,t_{last,i}\right)\right|\le {\delta }_i
\end{equation}

In (\ref{eq:2}), the measured value deviation increases from $v_i\left(t\right)$ to $v_i\left(t\right)+{\Delta }_i\left(t,t_{last,i}\right)$. If ${\Delta }_i\left(t,t_{last,i}\right)$ is assumed to have the uniform distribution with (\ref{eq:3}), then the variance of ${\Delta }_i\left(t,t_{last,i}\right)$ is $\frac{{(2 \times \delta)}^2_i}{12}$, which is added to the \textit{measurement noise covariance matrix}, R$\left(i,i\right)$, in the Kalman filter.
\newpage
\textbf{Algorithm for the SoD-based Kalman filter:} An algorithm is proposed here to appropriately improve the \textit{measurement update} part of the standard Kalman filter algorithm, which is adapted to the SoD event-generation condition by increasing the measurement noise covariance $\overline{R}_k$:
 \begin{enumerate}
 \item  Initialization set
 \begin{IEEEeqnarray}{ll}
 \hat{x}^-(0),{P}^-_0 \IEEEnonumber \\ 
 y_{last}=C\hat{x}^-\left(0\right)
 \end{IEEEeqnarray}
 \item  \textbf{Measurement update
 \begin{equation}
 \overline{R}_k=R
 \end{equation}
 if \textit{i}-th measurement data are received
 \begin{equation}
 \hat{y}_{last,i}=y_i\left(kT\right)
 \end{equation}
else
\begin{equation}
\overline{R}_k\left(i,i\right)=\overline{R}_k\left(i,i\right)+\frac{{(2 \times \delta)}^2_i}{12}
\end{equation}
end if
\begin{IEEEeqnarray}{lll}
K_k={P}^-_kC'(C{P}^-_kC'+\overline{R}_k)^{-1}\IEEEnonumber \\ 
\hat{x}\left(kT\right)=\hat{x}^-\left(kT\right)+K_k(\hat{y}_{last}-C\hat{x}^-(kT))\IEEEnonumber\\
P_k{=(I-K_kC)P}^-_k
\end{IEEEeqnarray}
}
\item  Project ahead
\begin{IEEEeqnarray}{lll}
\hat{x}^-\left((k+1)T\right)=\exp{\left(AT\right)}\hat{x}\left(kT\right)\IEEEnonumber\\
{P}^-_{k+1}=\exp{\left(AT\right)} P_k\exp{\left(A'T\right)}+Q_d
\end{IEEEeqnarray}
 \end{enumerate}
where $Q_d$ is the process noise covariance for the discretized dynamic system; $y_{last}$ is defined as (\ref{eq:jhjf}):
\begin{equation}\label{eq:jhjf}
y_{last}=[y_{last,1},y_{last,2},\dots ,y_{last,p}]'
\end{equation}

The presented event-triggered Kalman filter can also be employed to implement the distributed controllers and estimators in networked control systems. For further details on the development procedure and convergamce analysis, one can refer to \cite{Alavi2018a}. It should be noted that in the proposed event-triggered observer, convergence is obtained by using the Kalman optimal observer. However, choosing lower values of ${\delta }_i$ would result in the considerable reduction of the convergence time \cite{Li2016}.

\section{Signal Reconstruction Formulation and Estimator Update Rule}\label{reconstruction}
By using the SoD sampler, the input signal represents the time instants when the real signal has changed more than the specified $\delta$ value but also includes the fact that the signal stays in the region around the last sampled bounded by $\delta$. This additional information on the signal is considered as the implicit data in the sampled signal, which forms the optimization problem for signal reconstruction. Therefore, the samples provide the information of discrete-time equality constraints specified by the input signal, the additional implicit information makes of continuous time inequality constraints.

In this paper, the technique of Projection Onto Convex Sets (POCS) for bandwidth limited signal reconstruction from SoD samples, is proposed to optimally reconstruct the microgrid measurement signals with a low computation cost in real-time.  The POCS method was previously used for signal recovery from nonuniform samples \cite{Yeh1990}, and for image reconstruction from level crossings \cite{Zakhor1990, Yik-HingFung2006}. Send-on-Delta sampling is generalization of level-cross sampling, that considers the initial condition of the signal. This paper has extended the results of level-crossing sampling from \cite{Rzepka2018} to support send-on-delta sampling, which the readers are referred to for more details on convex optimization and projection algorithms.

\subsection{Implicit Information of Send-on-Delta Sampled Signal}
Send-on-Delta sampling is a type of event-based sampling, where each event shows a crossing of the signal $x(t)$ from a one dimensional region bounded by $\delta$ around the last sample. The event time instants $t_n \in \mathbb{Z}, n \in \mathbb{Z}$ are defined as:

\begin{equation}\label{eq_event}
t_n = \text{min}\{t > t_{n-1}, x(t) - x(t_{n-1}) > \delta\}
\end{equation}

The output of SoD sampler is the sequence of pairs $(t_n, x(t_n))$. The set of possible samples by assuming zero initial conditions is $ X_e = \{x(t_0), x(t_1), x(t_2),\dots,x(t_n)\}$. In order to formulate the convex optimization problem, a convex region for the possible range reconstructed signals is defined according to (\ref{eq_event}):

\begin{equation}
\theta^-(t) \le x(t) < \theta^+(t)
\end{equation}
where $\theta^-(t)$ and $\theta^-(t)$ are the piecewise constant lower and upper bound respectively created from the following constraints:

\begin{IEEEeqnarray}{lll}
\theta^-(t)=\{ r \in \mathbb{R}, r = x(k) - \delta, k \in t_n\} \IEEEnonumber \\
\theta^+(t)=\{ r \in \mathbb{R}, r = x(k) + \delta, k \in t_n\}
\end{IEEEeqnarray}

With this definition, the sign of the signal slope at the event instants ($t_n$)is defined as:

\begin{IEEEeqnarray}{lll}
S(t_n)=
\begin{cases}
x(t_n) - x(t_{n-1}),\ \ \ \ x(t_n) \neq x(t_{n-1}) \\
S(t_{n-1}),\ \ \ \ \ \ \ \ \ \ \ \ \ x(t_n) = x(t_{n-1})
\end{cases}
\end{IEEEeqnarray}

By using the previous definitions, the samples values along with the implicit information mathematically takes the form of sets membership. Therefore the solution for the reconstructed signal $x(t)$ will fall into the following convex sets ($C(\mathbb{R})$ denotes continuous function):

\begin{enumerate}
\item From the explicit information:
\begin{equation}
\xi = \{u(t) \in C(\mathbb{R}): u(t_n)=x(t_n) \text{ for all } n \in \mathbb{Z}\}
\end{equation}

\item From the implicit information:

\begin{equation}
\mathbb{I}= \{u(t) \in C(\mathbb{R}): \theta^- \le u(t) < \theta^+(t) \text{ for all } t \in \mathbb{R}\}
\end{equation}

\item From the knowledge that the signal is band-limited with maximum frequency $\Omega$
\begin{equation}
\mathbb{B} = \{u(t) \in \mathbb{L}^2(\mathbb{R}): \forall |w|
\end{equation}

\end{enumerate}

The set $\mathbb{B}$ is convex as the band-limited signals form a linear space. For the sets $\mathbb{I}$ and $\xi$, \cite{Rzepka2018} provides the proof of convexity. The reconstructed signal should be a member of the set $\xi \ \cap \ \mathbb{I} \ \cap \ \mathbb{B}$ as the constraint of the optimization. This constraint is usually wide that finding the optimal answer takes more computation. Fortunately, because $\theta^-(t) \le x(t) < \theta^+(t)$, one can easily derive that $\mathbb{I} \subset \xi$. Therefore, the constraint is limited to the boundary defined by $\mathbb{I} \ \cap \ \mathbb{B}$, which needs less computations for the task of real-time signal estimation.

\subsection{Projection onto Convex Sets Signal Reconstruction}
In order to solve the problem of POCS, two methods are proposed in literature, one-step projection and iterative projection. For the more detailed discussion of the mentioned methods, \cite{Rzepka2018} provides a good starting point. In this paper, as we are building a real-time event-based state estimator for microgrids, the later method of iterative projection onto convex sets is employed, which poses less computation with the price of losing a negligible precision. The basic idea behind iterative POCS is that by having two or more convex sets, on each iteration the initial solution is projected to one of them. Therefore, by iteratively repeating the projection to the sets, the initial guess gets closer to the optimal answer.

The projection of the a signal $g$ onto a continuous convex set $C$ will be another signal $\hat{x}(t)$ which is closest to signal $g$:

\begin{equation}
\hat{x} =  P_{Cg} = \text{arg} \min_{y \in C} ||g -y||
\end{equation}

where the projection $P_Cg$ is closer to any vector $ y \in C$ than g:

\begin{equation}
||P_{Cg} - x|| < ||g - y||
\end{equation}

For the event-based signal reconstruction problem, the initial guess $\hat{x}_0$ should be first projected onto convex set $\mathbb{B}$ with the following projection operator:

\begin{IEEEeqnarray}{lll}
P_{\mathbb{B}g}(t) = \hat{x}(t) \ast \frac{\Omega}{\pi} \text{sinc}(\Omega t) \IEEEnonumber \\ 
= \int_{-\infty}^{\infty} \hat{x}(\tau)\frac{\Omega}{\pi} \text{sinc}(\Omega (t - \tau))d\tau
\end{IEEEeqnarray}
having defined $sinc(y)=\frac{sin(y)}{y}$.

The projection operator onto convex set $\mathbb{I}$ for clipping the signal to bound defined by $\theta$ is:

\begin{IEEEeqnarray}{lll}
P_{\mathbb{I}g}(t)=
\begin{cases}
\theta^+(t),\ \ \ \  \hat{x}(t) > \theta^+(t) \\
\hat{x}(t),\ \ \ \ \ \ \theta^-(t) \le \hat{x}(t) < \theta^+(t) \\
\theta^-(t), \ \ \ \ \hat{x}(t) < \theta^-(t)
\end{cases}
\end{IEEEeqnarray}

Finally, by applying the operator for both projections, the desired accuracy of signal reconstruction will be achieved:
\begin{equation}
\hat{x}_{m+1} = P_{\mathbb{B}g}P_{\mathbb{I}g}\hat{x}_{m},\ \ \  m \in \mathbb{Z}
\end{equation}

The stopping condition for the number of iterations is application dependent, related to the accuracy needed for signal reconstruction. In this paper, experimentally we have chosen a value of 10 iterations, which provided a high accuracy.

\subsection{Estimator Update Rule of Mean-Square Error Comparator}\label{comparator}
The measurement signals from the sensors include levels of noise. Here in this paper, the noise type is considered as derivative of the Brownian motion (white or Gaussian noise). The event-triggered sampling of a signal with Gaussian noise generally leads to a non-Gaussian stochastic process, and therefore degrades the estimation accuracy and convergence of linear state estimators, such as the proposed event-triggered Kalman filter. Here we propose an estimator update rule that based the comparison of the reconstructed signal and the output of the Kalman filter, applies corrections to the measurements. The correction is an offset, that is added in the first stage of state estimator as described in the following:

\begin{IEEEeqnarray}{lll}
y_i\left(t_{last,i}\right)=
\begin{cases}
y_i(kT),\ \ \ \ ||y_{i_{predict}}- y_{i_{construct}}< \delta || \\
\\
y_{i_{construct}}(kT),\ \ ||y_{i_{predict}}- y_{i_{construct}} \geq \delta||
\end{cases}
\end{IEEEeqnarray}
where $y_{i_{predict}}$ and $y_{i_{construct}}$ are the output of the signal reconstructor and the event-triggered Kalman filter, respectively.
\section{Developed Setup for Estimator Validation}\label{prototype}

\begin{figure}[!t]
\centering
\includegraphics[width=3.5in]{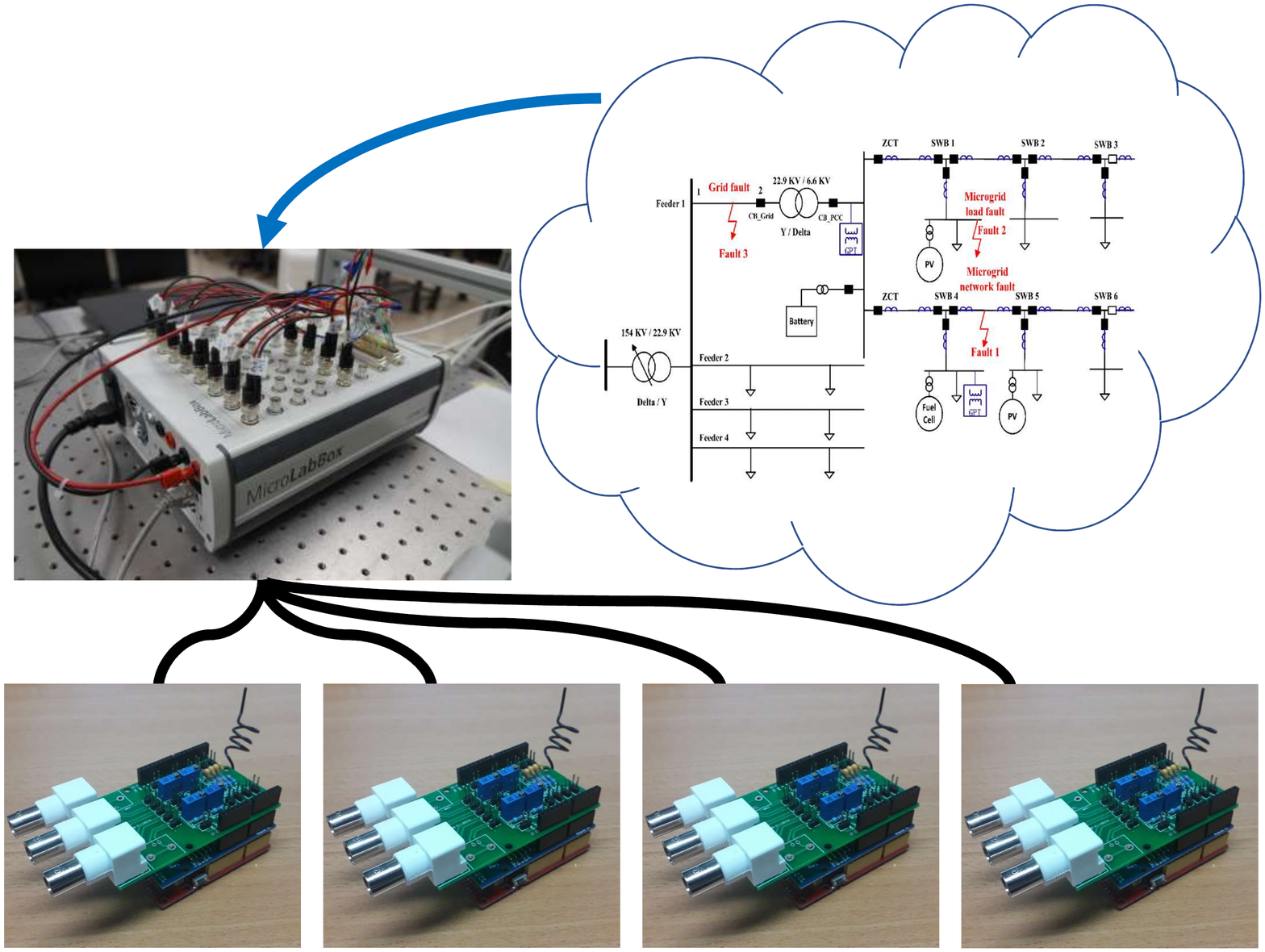}
\caption{Interface of measurement nodes with the real-time simulator.}
\label{hardware}
\end{figure}

In order to validate the proposed event-based state estimator, an IoT setup consists of several nodes supporting long range wide area network (LoRaWAN) communication protocol is designed using Seeeduino\textsuperscript{\textregistered} LoRaWAN nodes and a real-time microgrid simulator from dSPACE\textsuperscript{\textregistered} (Microlabbox DS1202). The nodes are connected to the real-time simulator via the BNC connectors that can be both Analog Outputs and Analog Inputs. The schematic of the setup is shown in \figurename ~\ref{hardware}. The real-time simulator allows the testing of different microgrid operation scenarios with only changing the simulation configuration in Matlab/Simulink software.

Since microgrids will be installed in private urban or rural areas, the monitoring software should be accessible easily by the operators, and also a well-designed human machine interface (HMI) is essential, in order to achieve the adequate situational awareness. In this work, the Thingsboard\textsuperscript{\textregistered} open-source software is used as the operator dashboard that shows the estimated state of the microgrid to the operator. Thingsboard is a web-based dashboard designer written in Java which provides different widgets to visualize the values received from the measurement nodes.

\begin{figure}[!t]
\centering
\includegraphics[width=3.5in,height=2.5in]{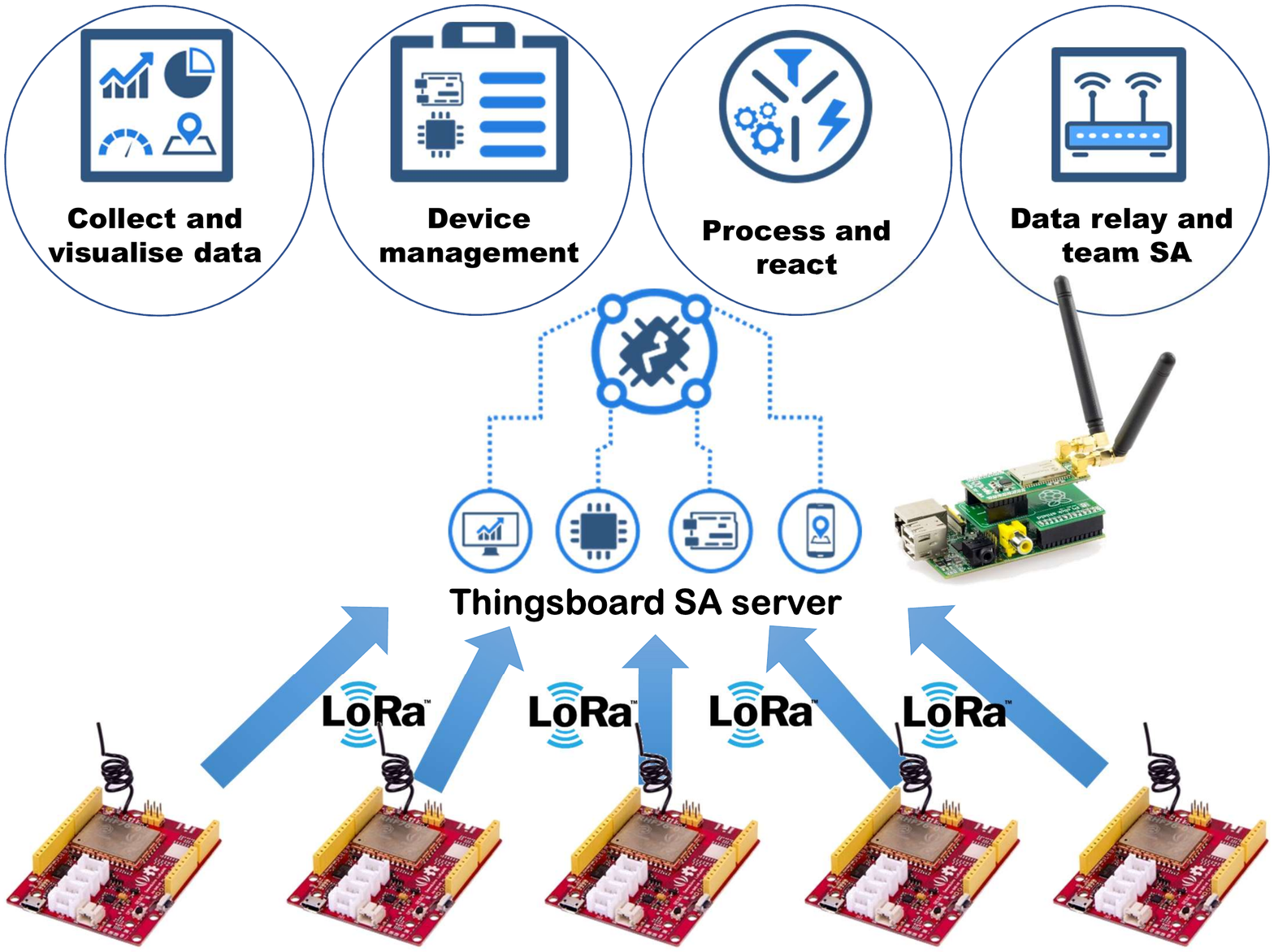}
\caption{Network architecture for microgrids based on the IoT protocols.}
\label{setup}
\end{figure}

The LoRaWAN protocol necessitates a gateway to be employed for the data collection and distribution. In this setup, a Raspberry Pi with the supporting communication module for the gateway operation is used. This gateway converts the received data from LoRaWAN nodes and transforms them into MQTT (Message Queuing Telemetry Transport) payloads which are transmitted to the MQTT broker. Thingsboard IoT software provides the MQTT broker which in this work is employed for data processing and archiving. The data collection architecture for the proposed  microgrid state estimator platform based on the IoT protocols is shown in \figurename ~\ref{setup}. By using the mentioned protocols and devices, the cost of monitoring of microgrid is considerably reduced. The developed hardware setup is comparably more affordable than the existing monitoring devices, which makes it an ideal choice for the big data collection and processing in smart grid.

The software stack developed for this device, fully supports the Arduino\textsuperscript{\textregistered} integrated development environment (IDE). Many libraries are developed for the Arduino that can be used seamlessly in this device. In addition, the battery life is extended due to the event-based communication. Hence, lower rating batteries can be used that leads to more cost reduction.

\section{Results and Discussion}\label{results}
In order to evaluate the proposed state estimation approach, in this section, an example microgrid model \cite{Alavi2018a} (canonical form), is simulated based on the developed data collection platform, defined as:
\begin{IEEEeqnarray}{lll}
\dot{x} = 
\left[\begin{matrix} 
0 & 1 & 0 & 0\\
0 & 0 & 1 & 0\\
0 & 0 & 0 & 1\\
-1 & -6 & -35.5 & -15
\end{matrix}\right]
x + w\IEEEnonumber \\
y=
\left[\begin{matrix} 
-2 & 4 & 0 & 3\\
0 & 10 & 0 & 1 
\end{matrix}\right]
x + v
\end{IEEEeqnarray}

The state of the microgrid is denoted by the vector $x(t)=[x_1(t)^T, x_2(t)^T, x_3(t)^T, x_4(t)^T]$, and the initial conditions are set as $x_0= [10, 3, -4, 5]$. The parameters of the proposed state estimator for the simulation are provided in Table \ref{tab:kalmanparamstbl}. The results are compared with the traditional Kalman filter and the superior performance of the proposed estimator is validated.

\begin{figure}[!h]
    \centering
	\subfloat{\includegraphics[width=4.3cm,height=4.3cm]{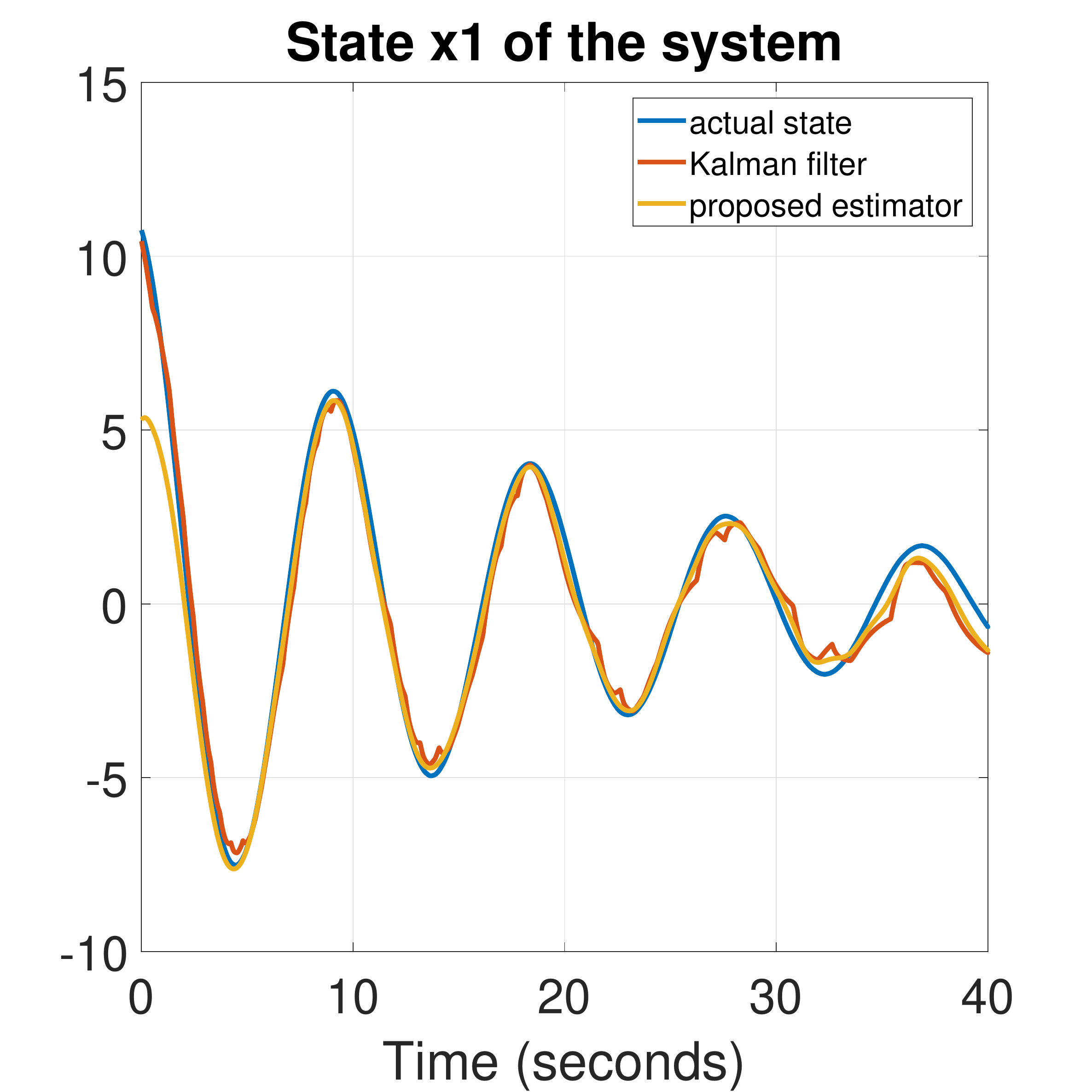}}
    \enspace
	\subfloat{\includegraphics[width=4.3cm,height=4.3cm]{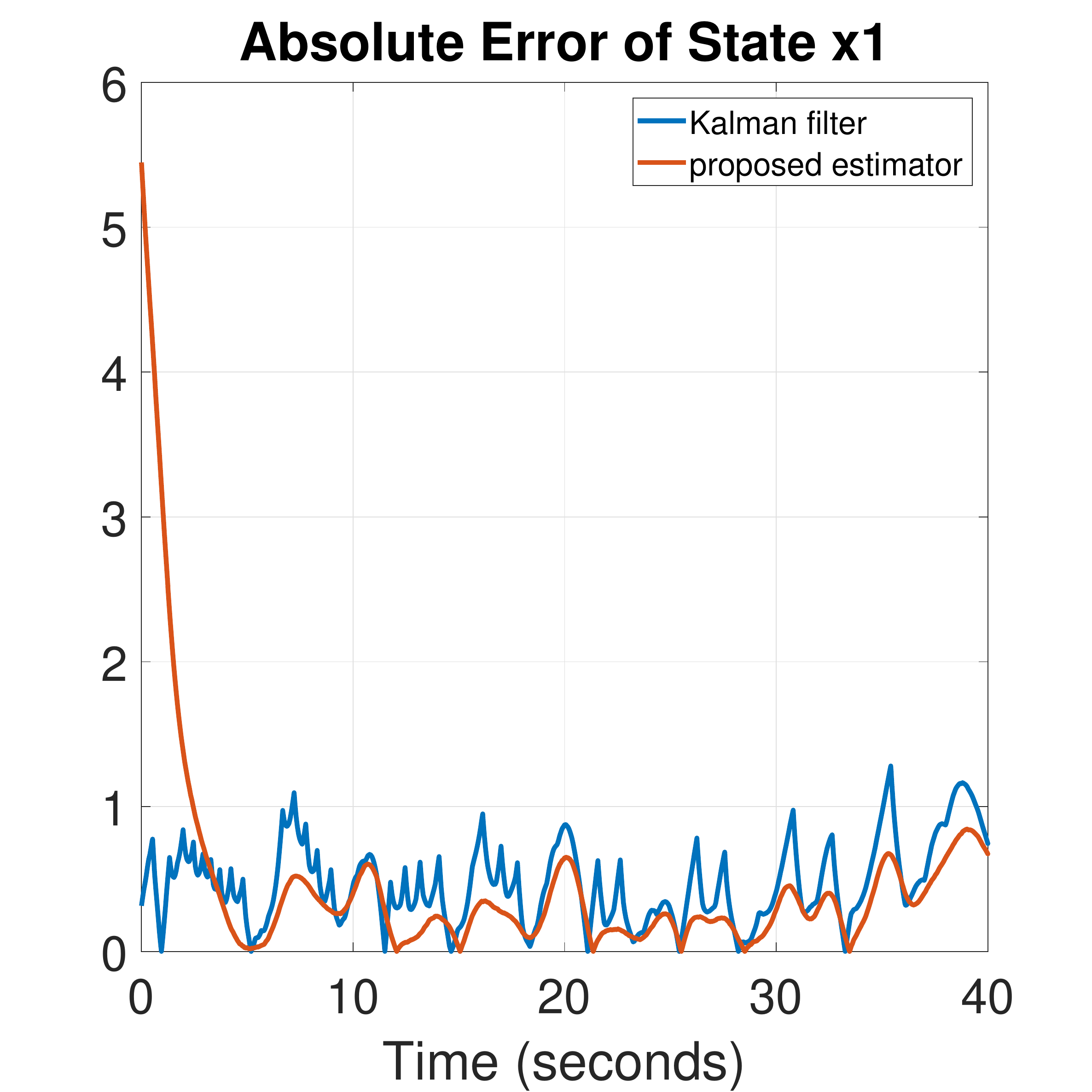}}
    \caption{Simulation result for state \textbf{x1} and error comparison}
    \label{fig:x1}
\end{figure}

\begin{figure}[!h]
    \centering
	\subfloat{\includegraphics[width=4.3cm,height=4.3cm]{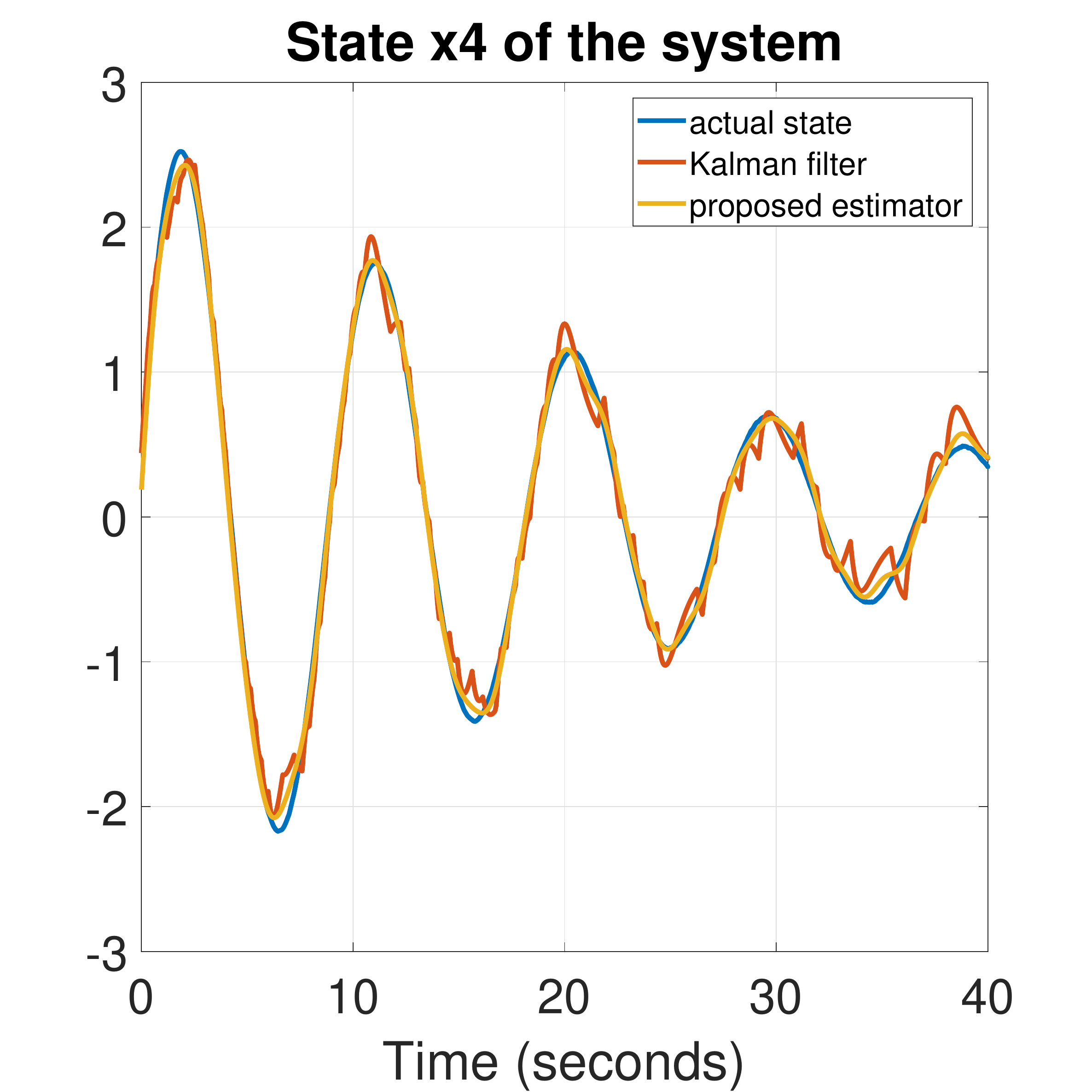}}
    \enspace
	\subfloat{\includegraphics[width=4.3cm,height=4.3cm]{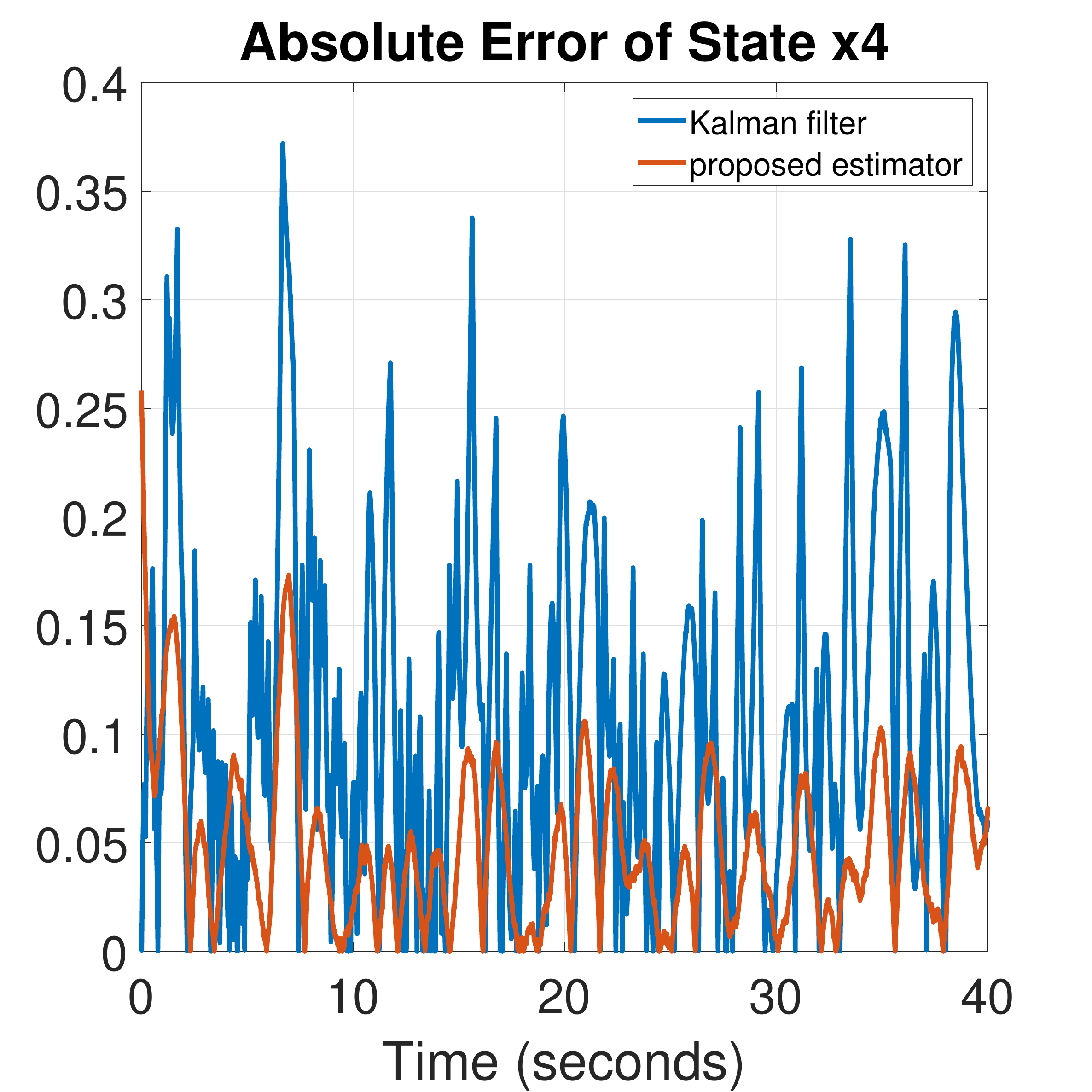}}
    \caption{Simulation result for state \textbf{x4} and error comparison}
    \label{fig:x4}
\end{figure}

The simulated system has two outputs and the number of events generated for each output based on SoD sampling is \underline{34} and \underline{84} events for a duration of \underline{40} seconds. This shows that with the small number of samples comparing to the time-triggered traditional Kalman filter, the estimator has achieved a better performance, as can be seen in the figures. \figurename ~\ref{fig:x1} to \figurename ~\ref{fig:x4} show the estimated state and the estimation error for both the proposed event-based estimator and the traditional Kalman filter. One can see that the estimation error is considerably lower. 

From the experimental point of view, there are limitations in the LoRaWAN communication network that may degrade the estimation accuracy. LoRaWAN protocol introduces a considerable delay of seconds to the transmission of the messages when the number of messages in a specific time, goes higher than the capability of the network. The number of message is related to the threshold $\delta$ of the SoD sampler, therefore a well designed tuning algorithm should be developed in order to relate the estimation error, SoD threshold and number of events. A delay compensated strategy would also solve this issue, which is part the future research in this paper.

\begin{table}[h]
\renewcommand{\arraystretch}{1.3}
\caption{Simulation parameters of the state estimator.}
\label{tab:kalmanparamstbl}
\centering
\begin{tabular}{|c|c|}
\hline
$\delta$ (SoD threshold) &  6 \\
\hline
$Q$ (Process Noise Covariance) & 0.1 \\
\hline
$R$ (Measurement Noise Covariance)& 0.36 \\
\hline
$T$ (Estimator Cycle Time) & 100 microseconds \\
\hline
\end{tabular}
\end{table}
\section{Conclusion}\label{conc}
This paper presented an optimal event-triggered state estimator for microgrids with the corresponding data collection architecture. A setup has been developed, which provides high performance data collection/estimation capabilities from smart meters. It has been shown that by using the developed estimation strategy, an adequate level of situational awareness can be achieved with lower installation and communication costs. Also the criteria for SoD sampling is justified, using event-based POCS signal reconstruction technique. In future, energy storages state of charge (SoC) will be also considered in the estimation problem, using the proposed technique.

\begin{figure}[!h]
    \centering
	\subfloat{\includegraphics[width=4.3cm,height=4.3cm]{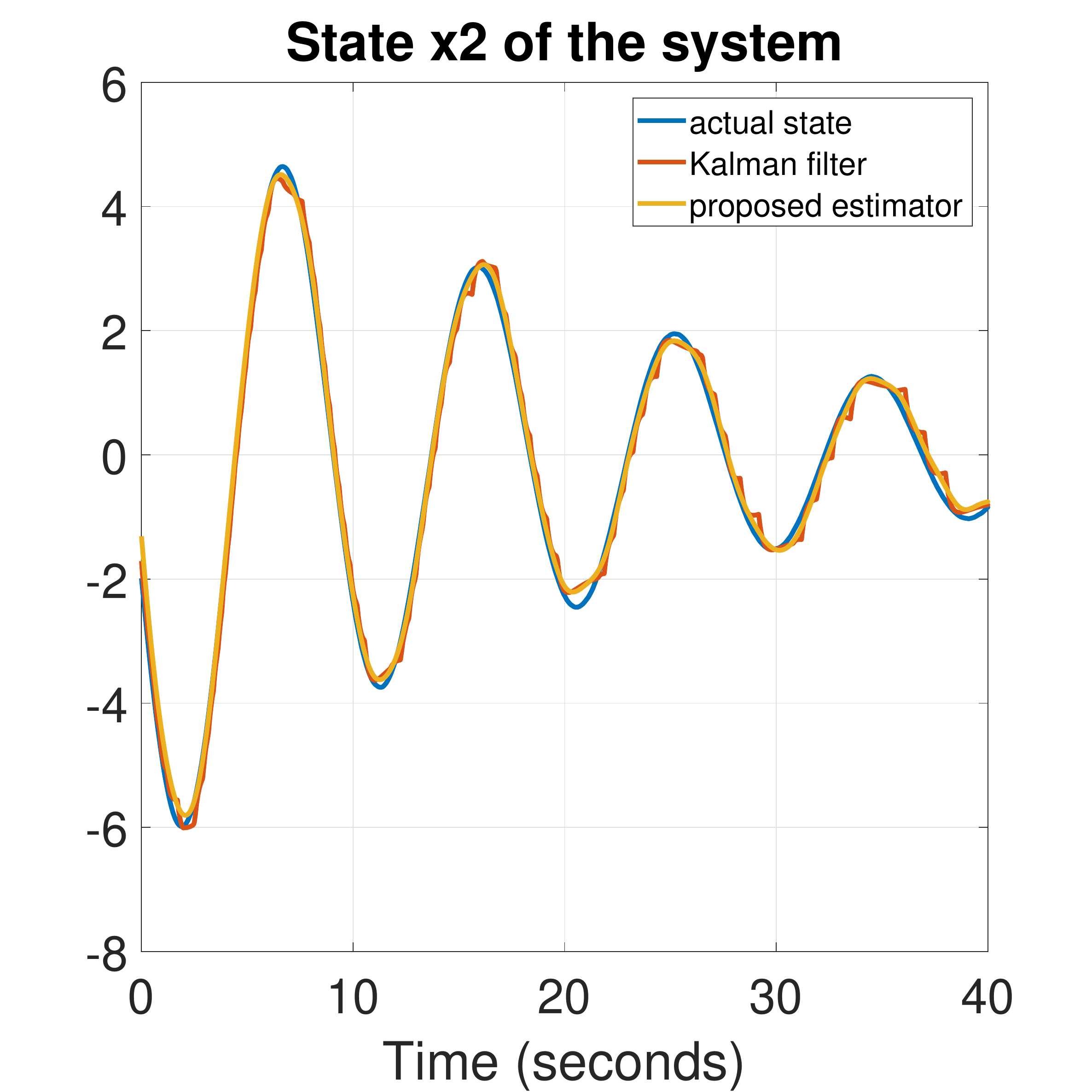}}
    \enspace
	\subfloat{\includegraphics[width=4.3cm,height=4.3cm]{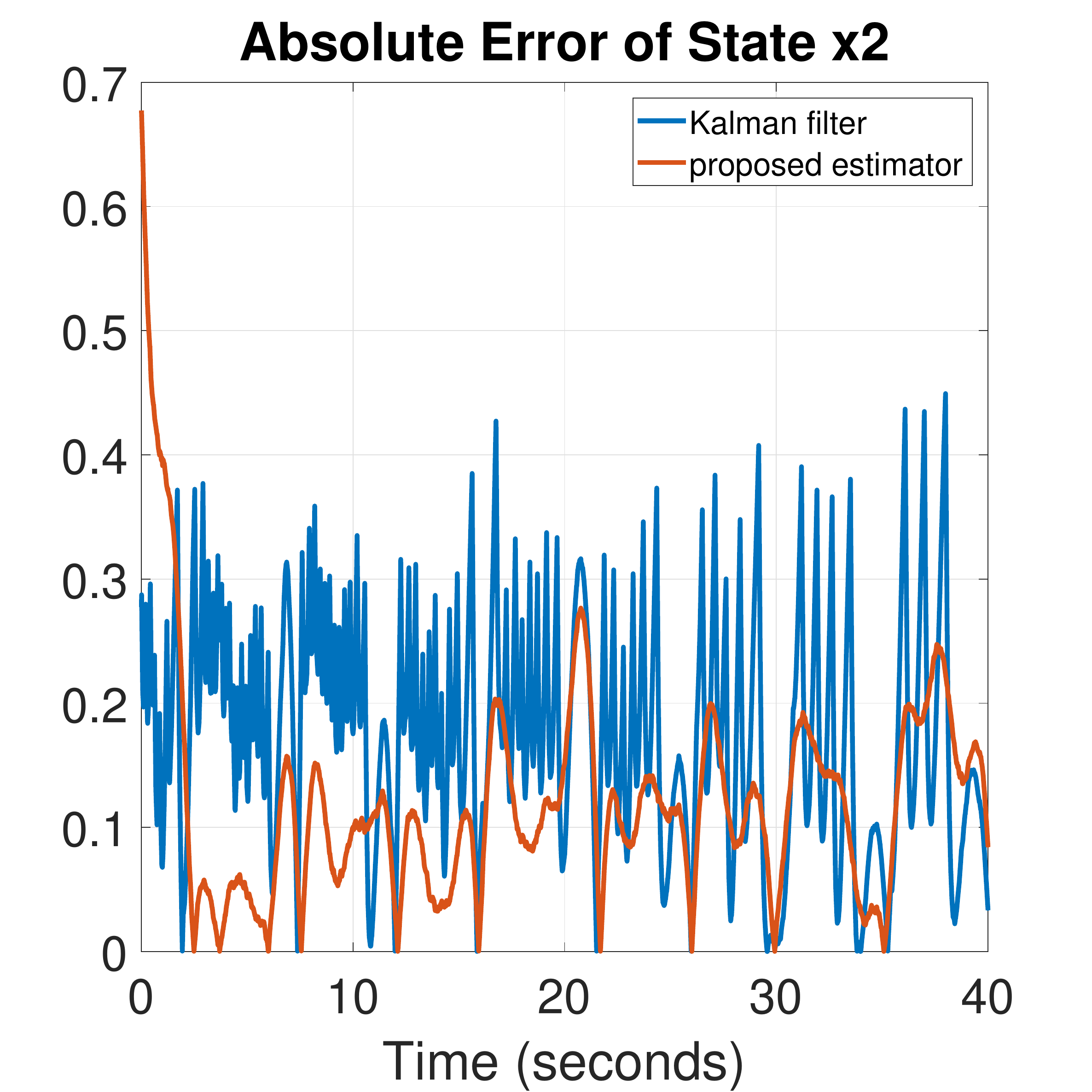}}
    \caption{Simulation result for state \textbf{x2} and error comparison}
    \label{fig:x2}
\end{figure}

\begin{figure}[!h]
    \centering
	\subfloat{\includegraphics[width=4.3cm,height=4.3cm]{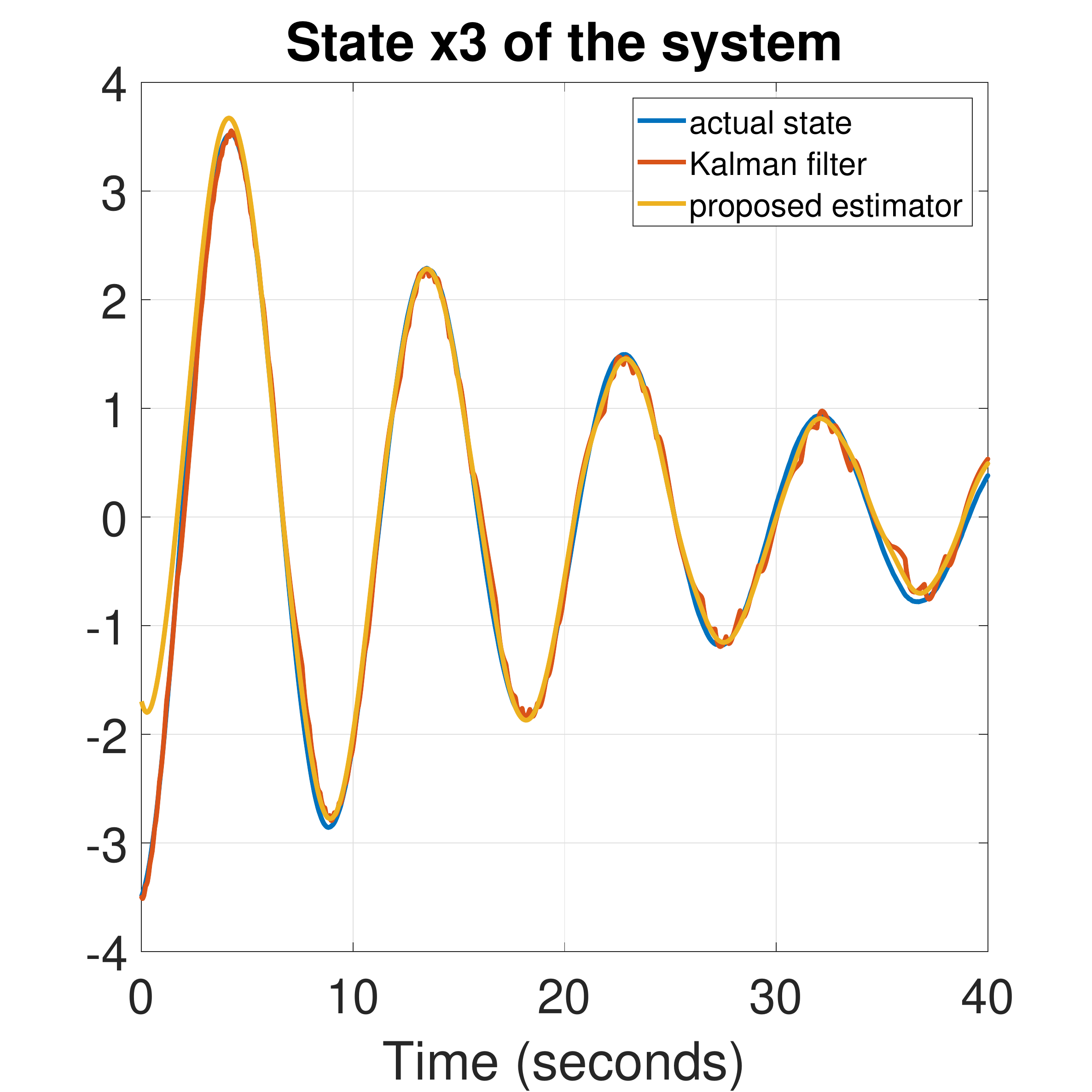}}
    \enspace
	\subfloat{\includegraphics[width=4.3cm,height=4.3cm]{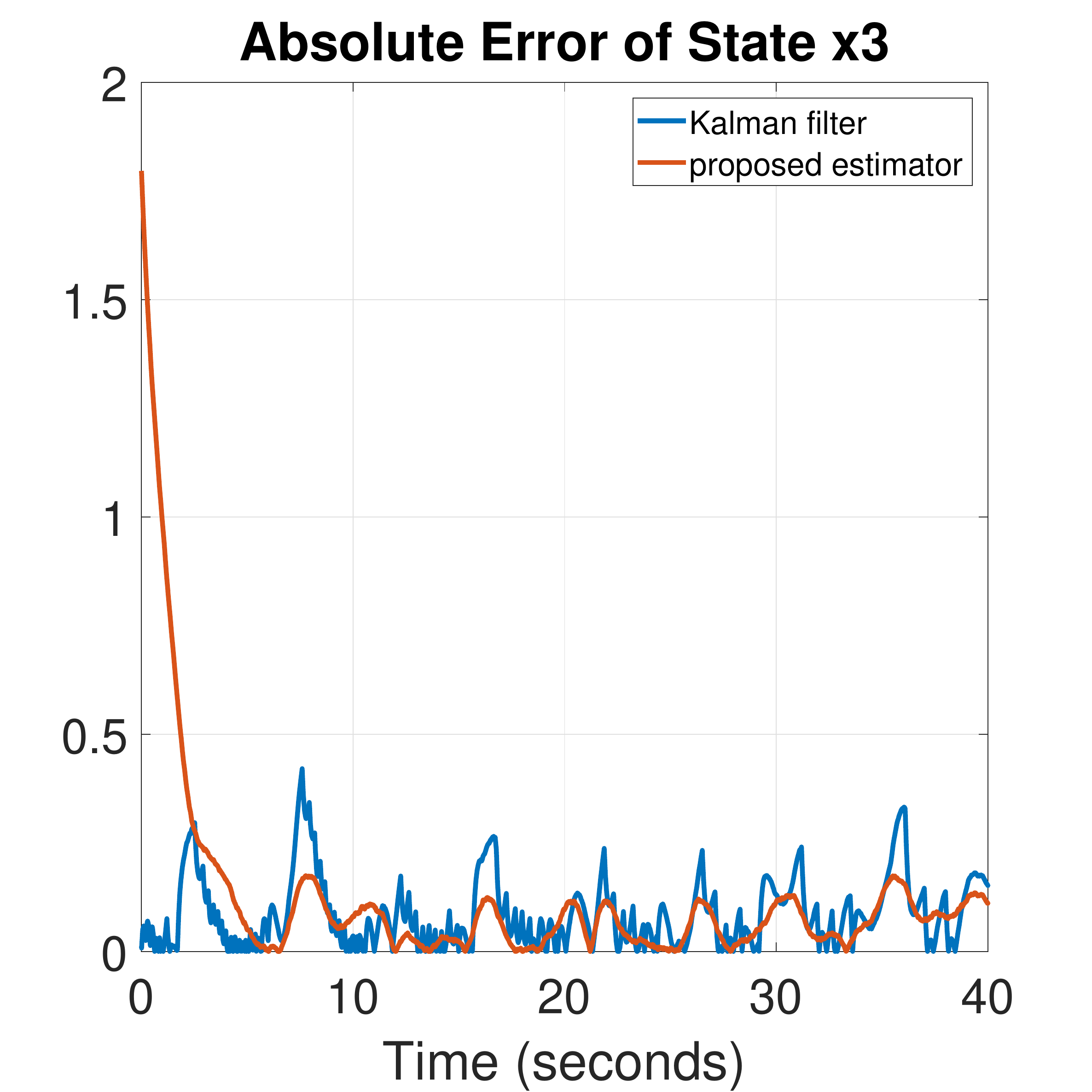}}
    \caption{Simulation result for state \textbf{x3} and error comparison}
    \label{fig:x3}
\end{figure}

\ifCLASSOPTIONcaptionsoff
  \newpage
\fi

\bibliographystyle{myIEEEtranbibstyle}

\bibliography{IEEEabrv,ref}

\end{document}